\documentclass{ws-ijmpa}

\begin{document}

\markboth{K.-I. Kondo}
{Magnetic condensation, Abelian dominance and instability of Savvidy vacuum}

\catchline{}{}{}{}{}
%
%%%%%%%%%%%%%%%%%%%%%%%%%%%%%%%%%%%%%%%%%%%%%%%%%%%%%%%%%%%%%%%%%%%%

\title{Magnetic condensation,     
Abelian dominance, \\
and instability of Savvidy vacuum  
in Yang-Mills theory
\footnote{This work is supported by 
Grant-in-Aid for Scientific Research (C)14540243 from Japan Society for the Promotion of Science (JSPS), 
and in part by Grant-in-Aid for Scientific Research on Priority Areas (B)13135203 from
the Ministry of Education, Culture, Sports, Science and Technology (MEXT).
}
}

\author{\footnotesize Kei-Ichi KONDO\footnote{
E-mail:  {\tt kondok@faculty.chiba-u.jp}.}}

\address{Department of Physics, Faculty of Science, 
Chiba University, Chiba 263-8522, Japan
%\footnote{State completely without abbreviations, the affiliation and mailing address, including country. Typeset in 8 pt italic.}
}

%\author{SECOND AUTHOR}

%\address{Group, Laboratory, Address\\
%City, State ZIP/Zone, Country
%}

\maketitle

%\pub{Received (Day Month Year)}{Revised (Day Month Year)}

\begin{abstract}
 We propose a novel type of color magnetic condensation  originating from  magnetic monopoles so that it provides the mass of off-diagonal gluons in the  Yang-Mills theory.  
 This dynamical mass generation enables us to explain the infrared Abelian dominance and monopole dominance by way of a non-Abelian Stokes theorem, which  supports the dual superconductivity picture of quark confinement.  
Moreover, we show that the instability of Savvidy vacuum disappears by sufficiently large color magnetic condensation.  
%In this derivation, the Cho--Faddeev--Niemi decomposition plays the important role.  

\keywords{magnetic condensation, Abelian dominance, monopole condensation, quark confinement, Savvidy vacuum.}
\end{abstract}

\section{Introduction}

In the Yang-Mills theory, i.e., non-Abelian gauge theory with asymptotic freedom,  Savvidy\cite{Savvidy77} has found that a nonperturbative vacuum with dynamically generated color magnetic field has lower vacuum energy density than the perturbative vacuum. Immediately after this discovery, however, Nielsen and Olesen (NO)\cite{NO78} have pointed out that the effective potential $V(H)$ of the color magnetic field $H$, when calculated explicitly at one-loop level, develops a pure imaginary part, although the real part agrees exactly with the Savvidy's result: The real part of $V(H)$ has an absolute minimum at $H=H_0$ away from $H=0$. The presence of the pure imaginary part implies that the Savvidy vacuum gets unstable due to gluon--antigluon pair annihilation. 
Since the energy eigenvalue $E_n$ of the massless {\it off-diagonal} gluons with the spin $S=1$ in the constant external magnetic field $H_z:=H_{12}$ is given by
\begin{equation}
 E_n^\pm=\sqrt{k_z^2+2gH_z(n+1/2) + 2gH_z S_z}$ $(n=0,1,2,\cdots) 
\end{equation}
where $S_z=\pm 1$,
  the instability is also understood as originating from the tachyon mode $n=0, S_z=-1$ (or the lowest Landau level with antiparallel spin to the external magnetic field), 
\begin{equation}
E_0^-=\sqrt{k_z^2-gH_z} .
\end{equation}

On the other hand, it is well known that in QED without asymptotic freedom, the non-zero magnetic field does not lower the vacuum energy and hence no magnetic condensation occurs, while the electric field causes electron--positron pair creation, destabilizing the QED vacuum. 

The NO instability of the Savvidy vacuum was derived based on the one-loop calculation of the effective potential.  Therefore,  some people consider it as indicating unreliability of the lowest-order loop calculation, i.e., artifact of the approximation.
% \cite{Kennaway04}.  
However, no one has demonstrated that the inclusion of higher order terms cures the instability. 
Moreover, the same problem exists also in the supersymmetric Yang-Mills theory in which the higher-order loop corrections are absent, since the covariantly constant background field strength is not supersymmetric.
% \cite{Kay83}. 

A way to circumvent the instability of the Savvidy vacuum is to introduce the magnetic domains with a finite extension into the vacuum,  in each of which the tachyon mode does not appear as far as  $k_z^2>gH_z$. 
This resolution  is called the Copenhagen vacuum. 
However, the Copenhagen vacuum breaks the Lorentz invariance and color invariance explicitly.

What type of vacuum is allowed and preferred in the Yang-Mills theory is an important question related to the physical picture of quark confinement.  
%Nevertheless, it is not yet full investigated so far. 
Can the instability be resolved in the one-loop level by a new mechanism? 
We have re-examined the stability of the Savvidy vacuum by incorporating new facts which have been obtained by the recent investigations related to quark confinement. 

\section{Facts}

First, it is useful to recall the assumptions taken in Nielsen and Olesen.\cite{NO78}

\begin{itemlist}
 \item 
The magnitude of the color magnetic field is uniform in spacetime and the  direction is specific $H_{12}=H_z$ (The direction is identified with the quantization axis of the off-diagonal gluon spin).

 \item 
The background gauge is taken as the gauge fixing condition.  
(Note that it is exactly the same as the Maximal Abelian gauge which has played the very important role in the recent investigations on quark confinement supporting  the dual superconductor picture.) 

\item 
 The off-diagonal gluons are treated as massless throughout the analysis. 
\end{itemlist}

In  order to re-examine the instability problem, it is important to recall the following facts which have been confirmed in the investigations on quark confinement since 1990.
\begin{itemlist}
 \item 
In the Maximal Abelian gauge, the infrared Abelian dominance \cite{tHooft81,EI82} and magnetic monopole dominance are discovered \cite{SY90} in the numerical simulations on the lattice.  
 \item 
This phenomena can be explained if the off-diagonal gluons acquire the mass $M$ which is much larger than the diagonal gluon mass.  
In fact, the off-diagonal gluon mass $M_X$ has been obtained in the numerical simulation on a lattice \cite{AS99,BCGMP03}, $M=1.2\rm{GeV}$.
See Ref.\cite{off-diagonal_mass,Kondo01} for analytical works. 
\end{itemlist}

We show that {\it the stability of the Savvidy vacuum is restored due to the dynamical mass generation of off-diagonal gluons  caused by a novel type of magnetic condensation coming from the magnetic monopoles}. 
This implies that quark confinement can be compatible with the stability of the Savvidy vacuum without resorting to the Copenhagen vacuum.
These are quite natural and  consistent results, since the condensation of magnetic monopoles is the key concept in a promising picture for understanding quark confinement, i.e.,  the dual superconductor picture.

As a technical device, we apply the Cho--Faddeev--Niemi (CFN)\cite{Cho80,FN98} decomposition  to SU(2) Yang-Mills theory to extract the magnetic monopole degrees of freedom explicitly from the non-Abelian gauge potential.

\section{Advantages of our method using the CFN decomposition}

\begin{itemlist}
 \item 
In our approach using the CFN decomposition, 
the direction  of the color magnetic field 
$\mathbb{H}_{\mu\nu}(x)=H_{\mu\nu}(x) \bm{n}(x)$ 
can be chosen arbitrary at every spacetime  point $x$ by using a unit vector $\bm{n}(x)$ indicating the color direction.
The Lorentz and color rotation symmetry is not broken by considering 
$
\|\mathbb{H}\| :=   \sqrt{\mathbb{H}_{\mu\nu} \cdot \mathbb{H}_{\mu\nu}} \equiv  g^{-1}   \sqrt{(\bm{n} \cdot (\partial_\mu \bm{n} \times \partial_\nu \bm{n}))^2}
$. 
It is invariant also under the color reflection, 
$\bm{n}(x) \rightarrow - \bm{n}(x)$.

 \item 
This formalism enables us to specify the physical origin of magnetic condensation as arising from the magnetic monopole 
through the relation, 
$
H_{\mu\nu}(x) := - g^{-1} \bm{n}(x) \cdot (\partial_\mu \bm{n}(x) \times \partial_\nu \bm{n}(x))
$.  
This is a microscopic description of the dynamically generated color magnetic field, in contrast to the Savvidy, Nielsen and Olesen.

\item
 The non-Abelian Wilson loop operator can be rewritten in terms of the CFN variables through the Diakonov--Petrov version of the non-Abelian Stokes theorem.\cite{DP89,KondoIV}  Hence we can separate the contribution from the magnetic variables in the Wilson loop average for examining the magnetic monopole dominance.

\item 
We can discuss the implications to the Faddeev-Skyrme model\cite{FN02} with knot soliton as a low-energy effective theory of Yang-Mills theory.    
This model is expected to describe  glueballs as knot solitons. 

\end{itemlist}

\section{CFN decomposition}

%%%%%%%%%%%%%%%%%%%%%%%%%%%%%%%%%%%%%%%%%%%%%%%%%%%%%%%%%%%%

\begin{figure}[htbp]
\begin{center}
\includegraphics[height=5cm]{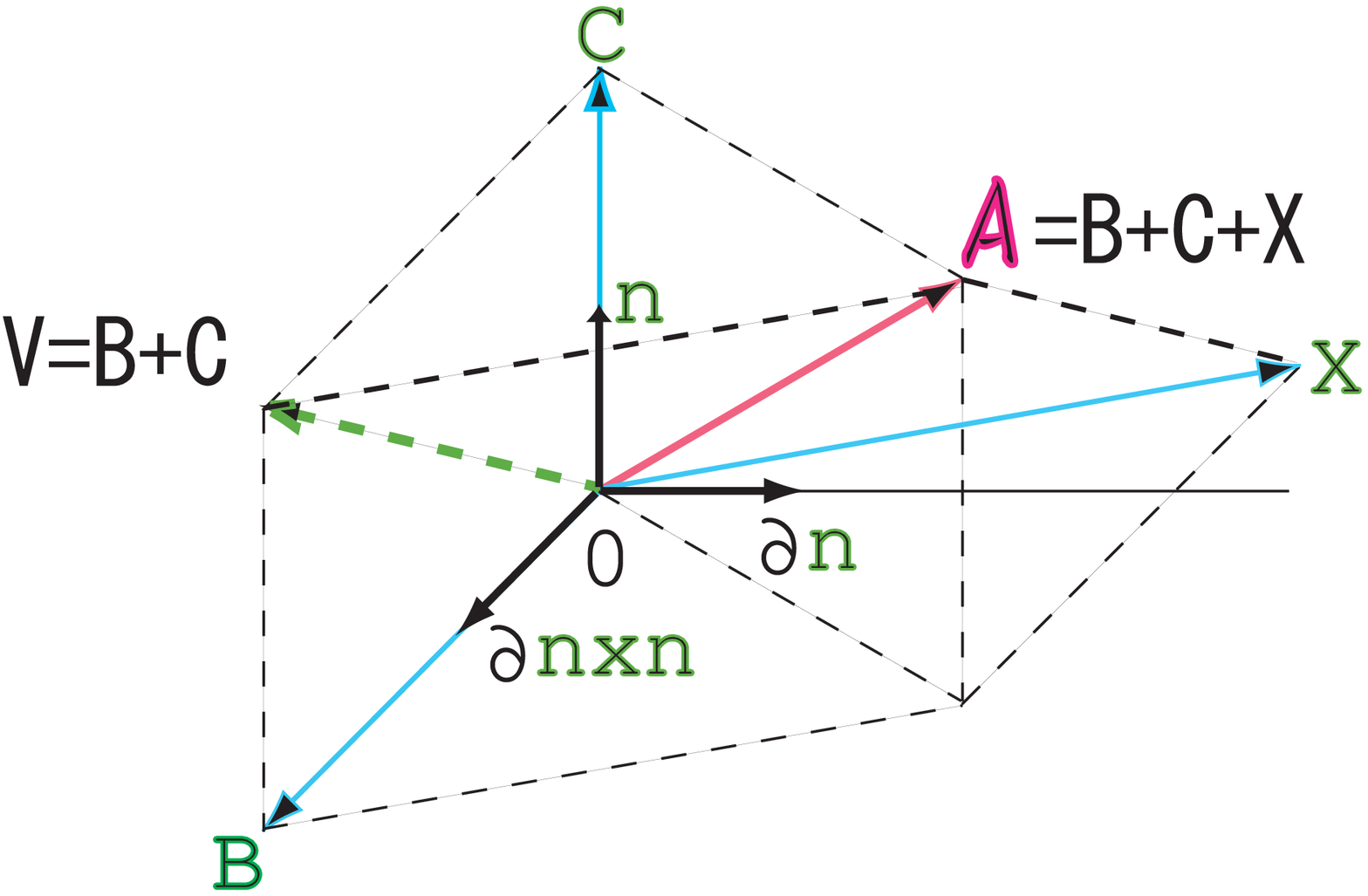}
\caption{\small The CFN decomposition of the gluon potential.}
\label{fig:CFNdecomp}
\end{center}
\end{figure}

%%%%%%%%%%%%%%%%%%%%%%%%%%%%%%%%%%%%%%%%%%%%%%%%%%%%%%%%%%%%

We adopt the Cho-Faddeev-Niemi (CFN) decomposition for the non-Abelian gauge field \cite{Cho80,FN98} for extracting the topological configurations explicitly, such as a magnetic monopole (of Wu-Yang type) and multi-instantons (of Witten type).   
%\rightline{[Cho,1980;1981] [Faddeev-Niemi, hep-th/9807069; hep-th/9812090; hep-th/9907180]}
By introducing  a three-component vector field $\bm{n}(x)$ with unit length, i.e., 
$\bm{n}(x) \cdot \bm{n}(x) = 1$, in the SU(2) Yang-Mills theory, 
the non-Abelian gauge field $\mathcal{A}_\mu(x)$ is decomposed as
\begin{eqnarray}
   \mathcal{A}_\mu(x) 
%=  (\mathcal{A}_\mu^A(x)) 
%\nonumber\\
%=& \underbrace{A_\mu(x) \bm{n}(x)}_{\mathbb{A}_\mu} 
% +  \underbrace{\left(  g^{-1} + \sigma(x) \right) \partial_\mu \bm{n}(x) \times \bm{n}(x)  + \rho(x) \partial_\mu \bm{n}(x)}_{\mathbb{W}_\mu} 
%\nonumber\\
%=& 
=\overbrace{ \underbrace{ C_\mu(x) \bm{n}(x)}_{\mathbb{C}_\mu(x)}
 +    \underbrace{ g^{-1}  \partial_\mu \bm{n}(x)  \times \bm{n}(x) }_{\mathbb{B}_\mu(x)}}^{\mathbb{V}_\mu(x)}  
 + \mathbb{X}_\mu (x) ,
\end{eqnarray}
where we have used the notation: 
$\mathbb{C}_\mu(x):=C_\mu(x) \bm{n}(x)$,  
$\mathbb{B}_\mu(x):=g^{-1} \partial_\mu \bm{n}(x) \times \bm{n}(x)$
and
$\mathbb{V}_\mu(x):=\mathbb{C}_\mu(x)+\mathbb{B}_\mu(x)$. 
By definition,  
$\mathbb{C}_\mu$ is parallel to $\bm{n}$, while $\mathbb{B}_\mu$ is orthogonal to $\bm{n}$.  We require $\mathbb{X}_\mu$ to be orthogonal to $\bm{n}$,  i.e., 
$\bm{n}(x) \cdot \mathbb{X}_\mu(x)=0$.  
We call $\mathbb{C}_\mu(x)$ the restricted potential, while $\mathbb{X}_\mu(x)$ is called the gauge-covariant potential and 
 $\mathbb{B}_\mu(x)$ can be identified with the non-Abelian magnetic potential. 
In the naive Abelian projection,  
$\mathbb{C}_\mu(x)$ corresponds to the diagonal component, 
while $\mathbb{X}_\mu(x)$ corresponds to the off-diagonal component, 
apart from the magnetic part $\mathbb{B}_\mu(x)$. 
Accordingly, the non-Abelian field strength $\mathcal{F}_{\mu\nu}(x)$ is decomposed as
\begin{eqnarray}
  \mathcal{F}_{\mu\nu} := \partial_\mu \mathcal{A}_\nu - \partial_\nu \mathcal{A}_\mu + g \mathcal{A}_\mu \times \mathcal{A}_\nu
= \mathbb{E}_{\mu\nu} + \mathbb{H}_{\mu\nu} + \hat{D}_\mu \mathbb{X}_\nu - \hat{D}_\nu \mathbb{X}_\mu + g \mathbb{X}_\mu \times \mathbb{X}_\nu , 
\end{eqnarray}
where we have introduced the covariant derivative, 
$
  \hat{D}_\mu[\mathbb{V}]  \equiv \hat{D}_\mu  := \partial_\mu   + g \mathbb{V}_\mu \times  ,
$
and defined the two kinds of {\it Abelian} field strength:
\begin{eqnarray}
  \mathbb{E}_{\mu\nu} &=& E_{\mu\nu} \bm{n}, \quad
E_{\mu\nu} := \partial_\mu C_\nu - \partial_\nu C_\mu ,
\\
  \mathbb{H}_{\mu\nu}  
&=& \partial_\mu \mathbb{B}_\nu - \partial_\nu \mathbb{B}_\mu + g \mathbb{B}_\mu \times \mathbb{B}_\nu
=  - g \mathbb{B}_\mu \times \mathbb{B}_\nu 
=  - g^{-1}   (\partial_\mu \bm{n} \times \partial_\nu \bm{n})   .
\end{eqnarray}
Here $\mathbb{H}_{\mu\nu}$ is the magnetic field strength proportional to $\bm{n}$ and is of the form
\begin{eqnarray}
  \mathbb{H}_{\mu\nu} =& H_{\mu\nu} \bm{n}, \quad
H_{\mu\nu} := - g^{-1} \bm{n} \cdot (\partial_\mu \bm{n} \times \partial_\nu \bm{n})  
= \partial_\mu h_\nu - \partial_\nu h_\mu ,
%=  \partial_\mu B_\nu - \partial_\nu B_\mu ,
\end{eqnarray}
since $H_{\mu\nu}$ is shown to be locally closed and it can be exact locally with the Abelian magnetic potential $h_\mu$. 
This is because  $H_{\mu\nu}$ represents the color magnetic field generated by magnetic monopoles.\cite{KondoII}

We adopt the Maximal Abelian gauge in the CFN decomposition,
$
 \bm{\chi} := D_\mu[\mathbb{V}] \mathbb{X}_\mu = 0 .
$ 
It is shown that the effective action to one-loop order is obtained by integrating out all the fields except for $\bm{n}$ and $C_\mu$ as
\begin{eqnarray}
 Z[J] &=& \int \mathcal{D}\bm{n}\delta(\bm{n} \cdot \bm{n}-1) 
\int \mathcal{D}C_\mu   
  e^{ - S_{EFF}   }  , 
\\
S_{EFF} &=& \int_{x} \frac{1}{4} [\mathbb{E}_{\mu\nu} + \mathbb{H}_{\mu\nu} ]^2
+  \frac{1}{2}\ln \det \{ (\mathbb{Q}){}_{\mu\nu} \} 
- \ln \det \{ -D_\mu[\mathbb{V}]D_\mu[\mathbb{V}] \}  
\nonumber\\ &&
+   \frac{1}{2} \ln \det \{ D_\mu[\mathbb{V}] (\mathbb{Q}^{-1}){}_{\mu\nu} D_\nu[\mathbb{V}] \} + \cdots ,
\\
 \mathbb{Q}_{\mu\nu}^{AB} 
 &:=&  -  (D_\rho[\mathbb{V}] D_\rho[\mathbb{V}])^{AB} \delta_{\mu\nu}  + g \epsilon^{ABC} n^C K_{\mu\nu}   ,
\quad 
 K_{\mu\nu} := 
 2(E_{\mu\nu} +  H_{\mu\nu} ) .
\label{defQ3}
\end{eqnarray}
In the one-loop level, the calculation of the effective potential reduces to the estimation of the logarithmic determinants.  

In the massless case,
the logarithmic determinants are calculated by using the $\zeta$-function regularization as 
\begin{eqnarray}
 & \frac{1}{2}\ln \det \{ (\mathbb{Q}){}_{\mu\nu} \} 
- \ln \det \{ -D_\mu[\mathbb{V}]D_\mu[\mathbb{V}] \}  
+   \frac{1}{2} \ln \det \{ D_\mu[\mathbb{V}] (\mathbb{Q}^{-1}){}_{\mu\nu} D_\nu[\mathbb{V}] \} 
%\nonumber\\
%=& \ln \det A_{+} + \ln \det A_{-} , 
%\quad
%  A_{\pm} := -(\partial_\mu + igh_\mu)^2 \pm 2g\|\mathbb{H}\| , 
\nonumber\\
 =& - \int d^4x {1 \over (4\pi)^2} \lim_{s \rightarrow 0} {d \over ds} {\mu^{2s} \over \Gamma(s)} \int_{0}^{\infty} d\tau \tau^{s-2} \  2g\|\mathbb{H}\| 
 \left[ {e^{-3g\|\mathbb{H}\| \tau} \over 1-e^{-2g\|\mathbb{H}\| \tau}} 
+ {e^{g\|\mathbb{H}\| \tau} \over 1-e^{-2g\|\mathbb{H}\| \tau}} \right] .
% \left[ {e^{-2g\|\mathbb{H}\| \tau} \over e^{+g\|\mathbb{H}\| \tau}-e^{-gH \tau}} 
%+ {e^{+2g\|\mathbb{H}\| \tau} \over e^{+gH \tau}-e^{-g\|\mathbb{H}\| \tau}} \right] .
\label{det1}
\end{eqnarray}
Then the effective potential of the magnetic field is obtained as\cite{Cho03}
\begin{eqnarray}
  V(\|\mathbb{H}\|) = {1 \over 4}\|\mathbb{H}\|^2 \left[ 1 + {b_0 \over (4\pi)^2}g^2 \left( \ln {g\|\mathbb{H}\| \over \mu^2} -c \right) \right] 
+ \rm{pure~imaginary~part} %\left({-i \over 8\pi}g^2H^2 \right) 
,
\label{effpot}
\end{eqnarray}
where $b_0={22 \over 3}$ and 
 $-b_0$ agrees with the first coefficient of the $\beta$-function, $\beta(g)={-b_0 \over 16\pi^2} g^3 + O(g^5)$.  
The pure imaginary part proportional to 
$g^2\|\mathbb{H}\|^2$
is the signal of instability pointed out by Nielsen and Olesen. 

In the massive case 
\begin{equation}
M_X^2 = g^2 \langle \mathbb{B}_\mu \cdot \mathbb{B}_\mu \rangle \not= 0 ,
\end{equation}
 the effective potential is modified into\cite{Kondo04}
\begin{eqnarray}
  V_r(\|\mathbb{H}\|) 
 &=&  {1 \over 4}\|\mathbb{H}\|^2 \Biggr[ 1 -  {8 \over (4\pi)^2}   g^2 \Biggr\{  \left[ 2 \zeta\left(-1, \frac{3+r}{2} \right) + r \right] 
 \ln  {g\|\mathbb{H}\| \over \mu^2}  
\nonumber\\ 
 &+&  \left[ 2 \zeta'\left(-1,\frac{3+r}{2} \right)   + \frac{-1+r}{2} \ln \frac{-1+r}{2} +  \frac{1+r}{2} \ln \frac{1+r}{2}  \right]  
 + c'  
\Biggr\} \Biggr] 
,
\label{det4}
\end{eqnarray}
where we introduced the  generalized $\zeta$ function $\zeta(z,\lambda)$ and the ratio defined by
\begin{eqnarray}
 r  :=M_X^2/(g\|\mathbb{H}\|) = g^2 \langle \mathbb{B}_\mu \cdot \mathbb{B}_\mu \rangle/(g \langle \|\mathbb{H}\|  \rangle)  .
\end{eqnarray}
$V_r(\|\mathbb{H}\|)$ is a real-valued function for $r \ge 1$, while it becomes complex-valued for $r<1$ and reduces to the NO result (\ref{effpot}) in the limit $r=0$. 
%See Ref.\cite{KMS04} for more details. 

\section{Conclusion}

We have discussed
\begin{romanlist}[(iii)]
 \item 
 The magnetic condensation of mass dimension 2 (spontaneous or dynamical generation of color magnetic field) can occur, i.e.,
$\langle \mathbb{B}_\mu \cdot \mathbb{B}_\mu \rangle>0$, 
$\langle \| \mathbb{H} \| \rangle:=\langle \sqrt{(\mathbb{B}_\mu \times \mathbb{B}_\nu)^2} \rangle>0$,  
as a consequence of gluonic interactions due to  magnetic monopole degrees of freedom which are extracted by the CFN decomposition and expressed through $\bm{n}$, i.e.,  
$
  g^2 \mathbb{B}_\mu \cdot \mathbb{B}_\mu =  (\partial_\rho \bm{n})^2   ,
$
and 
$    \sqrt{(g\mathbb{B}_\mu \times \mathbb{B}_\nu)^2}   
= g^{-1}   \sqrt{(\bm{n} \cdot (\partial_\mu \bm{n} \times \partial_\nu \bm{n}))^2}
\equiv   \|\mathbb{H}\|  . 
$
 \item 
  If a novel type of magnetic condensation occurs $\langle \mathbb{B}_\mu \cdot \mathbb{B}_\mu \rangle >0$, then the off-diagonal gluons acquire their mass $M_X$ as $M_X^2=g^2 \langle \mathbb{B}_\mu \cdot \mathbb{B}_\mu \rangle > 0$. 
This leads to the infrared Abelian dominance. This supports the dual superconductor picture for quark confinement.  The magnetic monopole dominance is also expected to hold. 

 \item 
 If the off-diagonal gluon mass $M_X$ obtained in this way is sufficiently large so that $M_X^2 \ (=g^2 \langle \mathbb{B}_\mu \cdot \mathbb{B}_\mu \rangle) > \langle g \|\mathbb{H}\|  \rangle$, the tachyon mode in the Savvidy vacuum is eliminated and the stability is restored. 
\end{romanlist}

In fact, the above statements are confirmed as follows. 
Even in the massive case, the existence of a magnetic condensation has been shown 
$
  \langle g \|\mathbb{H}\|  \rangle > 0 ,
$
within the one-loop level (improved by the renormalization group)\cite{Kondo04}.
The existence of another magnetic condensation,  
$
 g^2 \langle \mathbb{B}_\mu \cdot \mathbb{B}_\mu \rangle  >0 ,
$
can be shown\cite{Kondo04}  based on a simple mathematical identity 
$
 (\mathbb{B}_\mu \times \mathbb{B}_\nu) \cdot (\mathbb{B}_\mu \times \mathbb{B}_\nu)
=  (\mathbb{B}_\mu \cdot \mathbb{B}_\mu)^2 - (\mathbb{B}_\mu \cdot \mathbb{B}_\nu) (\mathbb{B}_\mu \cdot \mathbb{B}_\nu) , 
$
which yields a bound on $\mathbb{B}_\mu \cdot \mathbb{B}_\mu$,
$
  (\mathbb{B}_\mu \cdot \mathbb{B}_\mu)^2 - (\mathbb{B}_\mu \times \mathbb{B}_\nu) \cdot (\mathbb{B}_\mu \times \mathbb{B}_\nu) = (\mathbb{B}_\mu \cdot \mathbb{B}_\nu) (\mathbb{B}_\mu \cdot \mathbb{B}_\nu) \ge 0, 
$
i.e., 
$
 g^2 \langle \mathbb{B}_\mu \cdot \mathbb{B}_\mu \rangle \ge \langle g \|\mathbb{H}\|  \rangle >0 ,
$
leading to 
$r \ge 1$. 
Then the tachyon mode is removed. 
(The possible zero  mode can not be excluded.) 
A stronger bound is obtained\cite{Kondo04} by using the Faddeev--Niemi variable\cite{FN02}, 
$
 g^2 \langle \mathbb{B}_\mu \cdot \mathbb{B}_\mu \rangle \ge \sqrt{2} \langle g \|\mathbb{H}\|  \rangle ,
$
which yields the ratio,  
$
 r   \ge \sqrt{2} .
$
Thus the tachyon mode and the zero mode are removed.

The topics to be tackled in the future are 
\begin{itemlist}
 \item 
 Clarifying the relationship between magnetic condensation and magnetic monopole condensation, in connection to 
magnetic monopole dominance 
 \item 
 Extension to  the gauge group SU(N)

\item 
 Going beyond one-loop

\item
 Inclusion of quark (enables us to discuss the chiral symmetry breaking)

\end{itemlist}

\end{document}